\documentstyle[aps,twocolumn,prl]{revtex}
\begin{document}
\draft
\twocolumn[\hsize\textwidth\columnwidth\hsize\csname @twocolumnfalse\endcsname
\title{Intrinsic Decoherence in Mesoscopic Systems}
\author{P. Mohanty,
        E. M. Q. Jariwala, 
        and R. A. Webb     \\
        Center for Superconductivity Research,\\
        Department of Physics, University of Maryland, College Park, 
	Maryland 20742}

\maketitle

\begin{abstract}

We present measurements of the phase coherence time $\tau_\phi$ in six 
quasi-1D Au wires and clearly show that $\tau_\phi$ is temperature
independent at low temperatures. We suggest that zero-point fluctuations
of the intrinsic electromagnetic environment are responsible for the observed
saturation of $\tau_\phi$. We introduce a new functional form for the
temperature dependence and present the results of a calculation for the 
saturation value of $\tau_\phi$. This explains the observed temperature
dependence of our samples as well as  many 1D and 2D systems reported 
to date.

\end{abstract}
\pacs{PACS numbers: 03.65.Bz, 72.70.+m, 73.20.Fz, 73.23.-b}]  

\makeatletter
\global\@specialpagefalse
\let\@evenhead\@oddhead
\makeatother

\par

Perhaps the most fundamental property of a particle in any quantum system
is the time over which the phase coherence is maintained in its wave
function. It is well understood that coupling the quantum system to an
environment\cite{feynman} can cause a reduction in the constructive 
interferences of all possible Feynman paths due to changes induced in 
the environment by the particle and/or by phase randomization in the 
particle's wavefunction caused by the environment. In condensed matter
electron systems, the components of the environment which can cause
decoherence are the electron-phonon(EP), the electron-electron(EE), 
and magnetic impurity interactions\cite{chakravarty}. In addition,
under certain conditions dephasing can occur in the absence of 
any inelastic process\cite{stern}.
Elastic scatterings
from non-magnetic impurities are known not to cause phase 
randomization\cite{landauer}. The standard approach for determining the
phase coherence time $\tau_\phi$ in diffusive 1D and 2D systems is to fit
weak localization theory to the measured change in resistance as a 
function of magnetic field near zero field\cite{chakravarty,altshuler}.
Theoretically, this measured $\tau_\phi(T)$ should increase with 
decreasing temperature becoming infinite in very large systems at $T=0$
because both the EP and EE interactions produce a $\tau_\phi \sim 1/T^p$
where $p$ varies between $0.5$ and
$3$\cite{chakravarty,altshuler,stern,altshuler:book}.
Yet in every published experiment performed down to very
low temperatures, the phase coherence time is universally found to 
approach a temperature independent and
finite value\cite{giordano2:aupd,pepper2:si,hiramoto1:gaas,bergmann:au}.
The temperature at which this 
saturation occurs varies by orders of magnitude ranging from 10 K in
some GaAs devices to as low as 20 mK in a 2D Au film.

\par

There have been many different theoretical approaches aimed at     
understanding the temperature dependence of 
$\tau_\phi$\cite{chakravarty,stern,altshuler:book}.
At temperatures below which the phonons are 
important, Altshuler {\it et al.}\cite{altshuler:book} have shown that 
the EE process with large energy transfer should dominate with $\tau_\phi 
= \tau_{ee} \propto L_T$, where $L_T=\sqrt{\hbar D/k_BT}$, and $D$ is the
classical diffusion coefficient in $d$ dimensions. Subsequently, it was
suggested that at low temperatures an EE process with small energy transfer
will dominate the temperature dependence of the scattering rate with 
$\tau_\phi = \tau_{N} \propto 1/T^{2/3}$ \cite{altshuler:book}. This latter 
form has been verified by several papers where the Aharonov-Bohm phase of
the electron wavefunction is used to compute the mean square value of 
$\tau_\phi$, which is then related to the resistance of the phase coherent
volume of the system using the fluctuation-dissipation 
theorem\cite{chakravarty,stern}. Also many experiments have claimed to
find agreement with this form over some temperature 
range\cite{giordano2:aupd,pepper2:si}, but at lower temperatures they 
observe a much weaker temperature dependence or an approach to a complete
saturation in $\tau_\phi$.  

\par

In this letter we report an extensive set of experiments designed to 
understand what sample parameters control the magnitude and functional
form of the temperature dependence of  $\tau_\phi$. We also observe the 
saturation in $\tau_\phi$ at low temperatures, and show that it is not
due to heating or magnetic impurities as frequently suggested. We propose
that zero-point fluctuations of phase coherent electrons are responsible
for the observed saturation. This is fundamentally different from         
previous attempts to include intrinsic fluctuations in mesoscopic systems
where only the zero-point motion of impurity ions was             
considered\cite{kumar}.
We show that all our data can be surprisingly 
fit by one universal form that includes the zero-point decoherence time. 
This form fits most 1D and 2D data reported over the last 15 years. We 
present the results of the first calculation of this decoherence time.

\par

All our samples were fabricated from pure gold containing less than 1 ppm
of magnetic impurities and patterned into wires whose width, thickness, 
length and disorder varied considerably as shown in Table 1. A typical 
weak localization measurement for a 1D Au wire at 11 mK is shown in the 
inset to Fig. 1. We fit this data to the standard 1D form including both
the singlet and triplet terms to obtain the phase coherence length
$L_\phi$\cite{chakravarty}. The  phase coherence time is obtained by 
$\tau_\phi = L_\phi^2/D$. Our samples are 1D with respect to $L_\phi$ and
$L_T$, but most are 3D with respect to the elastic mean free path $l_e$.
Fig. 1 displays $\tau_\phi(T)$ from 11 mK to 7 K for one of our samples.
Below 200 mK the temperature dependence of $\tau_\phi$ is slower than 
expected from any theory\cite{altshuler}, and clearly saturates below 
40 mK. At temperatures above 2 K, the 
%
\begin{figure}
 \vbox to 7cm {\vss\hbox to 7cm
 {\hss\
   {\includegraphics{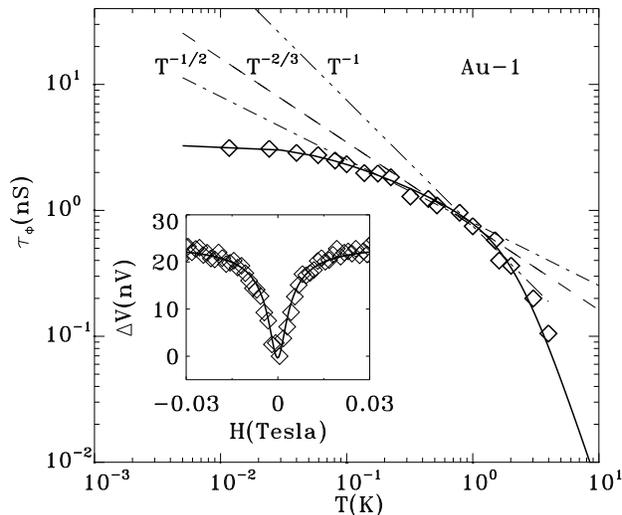}
   }
  \hss}
 }
\caption{Temperature dependence of $\tau_\phi$ for sample Au-1.
The broken lines are the functional forms expected from previous theories. The
solid line is a fit to Eq. (1) with phonons. The inset shows the typical weak
localization data taken with 2 nA at 11 mK with a fit to the standard 1D theory.}
\end{figure}
\noindent
EP interaction begins to dominate
the temperature dependence of $\tau_\phi$. In this paper we will assume
that the scattering time due to phonons is given by
$\tau_{ep} = A_{ep}/T^p$
with $p=3$ which is the most easily justified form\cite{altshuler:book}.
We have measured over twelve different 1D Au wires with
varying width from
35-210 nm, thickness from 20-135 nm, and length from 19-4120 $\mu$m.  
$D$ varied by more than a factor of 250, ranging from 0.00027 to 0.07 
m$^2$/s. Our low temperature $L_\phi$ varied by more than a factor of 40, 
from 0.43-18 $\mu$m, increasing with increasing $D$. Fig. 2 displays the 
temperature dependence of $\tau_\phi(T)$ for four of our samples representing
a broad range of disorder. For our samples, $\tau_\phi(T)>\hbar/k_BT$, the 
thermal time, over the entire data range. At low temperatures, all our 
$\tau_\phi(T)$ data show saturation or a much weaker temperature dependence 
than predicted from any theory. The temperatures at which these samples begin
to deviate from the theoretically expected form differ from one another, 
but they all show the same qualitative behavior.

\par

We believe that heating is not responsible for the observed saturation for
three reasons. Our experiments were done in the regime where $\tau_\phi$
is independent of current. In addition, we have measured the temperature
dependence of the change in resistivity $\Delta\rho_{ee}$ due to the EE 
interaction in a finite magnetic field sufficient to destroy weak 
localization. As shown in the inset to Fig. 2 we find that the correction
$\Delta\rho$ still remains temperature dependent down to our lowest 
temperatures, even though $\tau_\phi$ clearly saturates. The straight line 
is the theoretically expected behavior, 
$\Delta\rho_{ee}=(2e^2R^2wt/hL^2)\sqrt{\hbar D/k_BT}$\cite{altshuler}, 
demonstrating that the electrons are still in good contact with the thermal
bath. Finally, we have placed some of our samples in a second 
\begin{figure}
 \vbox to 7cm {\vss\hbox to 7cm
 {\hss\
   {\includegraphics{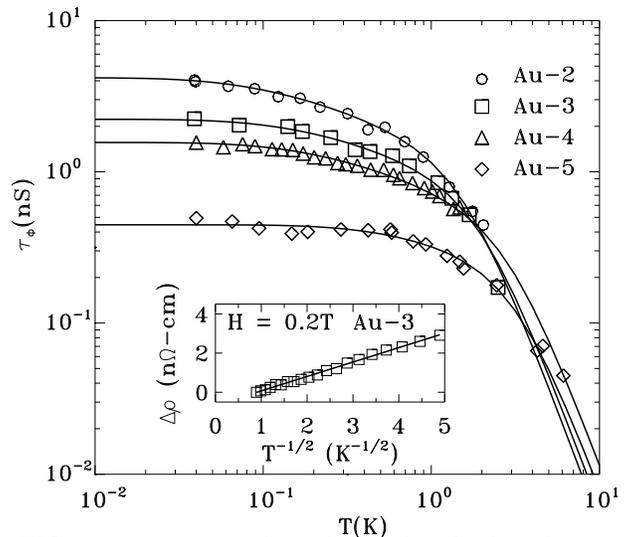}
   }
  \hss}
 }
\caption{Temperature dependence of $\tau_\phi$ for four
Au wires. Solid lines are fits to Eq. (1) with phonons. The
inset is the EE contribution to $\Delta\rho$
with the theoretical prediction.}
\end{figure}
\noindent
dilution
refrigerator with
higher levels of shielding of the external electromagnetic
environment at the sample site and find exactly the same $\tau_\phi(T)$.

\par

We have also  studied the effects of magnetic impurities on $\tau_\phi(T)$
by ion implanting  Fe, the dominant magnetic
impurity in Au, after
measuring the full $\tau_\phi(T)$ in the
clean sample. In Fig. 3, we 
compare the temperature dependence of $\tau_\phi$ for one sample before and
after implantation of $\sim$ 2.8 ppm of Fe. The effect of adding magnetic 
impurities is to lower the magnitude of the phase coherence time, but 
{\it not} to cause saturation in $\tau_\phi(T)$. The low temperature data 
is clearly temperature dependent in agreement with previous 
experiments\cite{bergmann:aufe,chris:cucr}. In addition,
the saturation of $\tau_\phi$ found in experiments on semiconductor
wires\cite{pepper2:si,hiramoto1:gaas} cannot be due to magnetic impurities
since these structures are thought not to contain any of these
impurities. The inset to Fig. 3 shows the
low temperature behavior of $\Delta\rho(T)$ for the implanted sample after
subtracting the EE contribution $\Delta\rho_{ee}$ determined from the clean 
sample before implantation. The straight line is the expected behavior 
$\Delta\rho \sim {\log}T$ for an AuFe Kondo system\cite{venkat}, containing
4.8 ppm of Fe\cite{daybell}.

\par

For the reasons given above, we are confident that the saturation in 
$\tau_\phi$ observed for all our clean Au samples represents a fundamental
quantum mechanical effect. We believe that the origin of the observed 
saturation in $\tau_\phi$ is that the zero-point fluctuations of the phase
coherent electrons are playing an important role in the dephasing process.
It is predicted that at low temperatures the mean square voltage in an 
electrical resistor will be finite at $T=0$ due to the zero-point 
fluctuations of the electrons\cite{feynman}. We propose that zero-point
fluctuations of the intrinsic electromagnetic 
environment\cite{landau} seen by the
phase coherent electrons should cause intrinsic dephasing and lead to a 
finite temperature saturation  
\begin{figure}
 \vbox to 7cm {\vss\hbox to 7cm
 {\hss\
   {\includegraphics{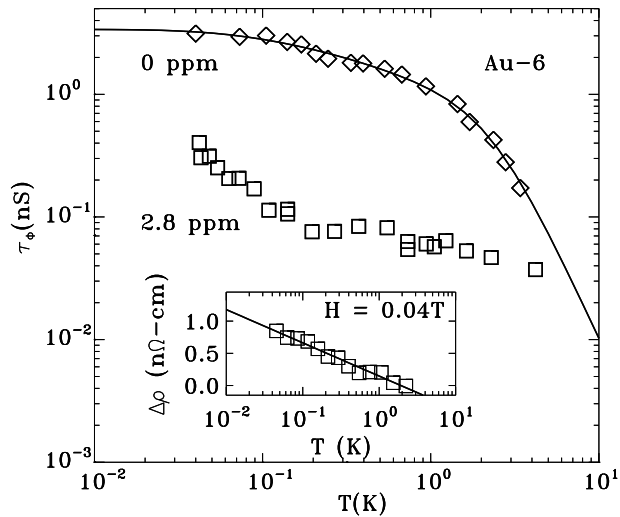}
   }
  \hss}
               }
\caption{Temperature dependence of $\tau_\phi$
        before (diamonds) and after (boxes) Fe implantation.
        The solid line is a fit to Eq. (1) with phonons. The inset 
        shows the $\log (T)$ dependence of $\Delta\rho$
        due to magnetic impurities with a theoretical fit.}
\end{figure}
\noindent  
of $\tau_\phi$. We have discovered that, at
low temperatures, one very simple form fits the temperature dependence of
$\tau_\phi(T)$ for all our Au samples,
\begin{eqnarray}
\tau_{\phi} = \tau_0{\tanh}\Big[{{\hbar \alpha \pi^2 D}\over {k_BT
L_T L_\phi^{0}}
}\Big]
=\tau_0 {\tanh}\Big[\alpha\pi^2 \sqrt{\hbar\over{\tau_0 k_B T}}\,\Big],
\end{eqnarray}
where $D=v_F l_e/d$, $\tau_0$ is the measured saturation value, and
$L_\phi^{0}= \sqrt{D\tau_0}$. Eq. (1) could have been anticipated
once the connection to the fluctuation-dissipation theorem\cite{landau} 
has been made because the inverse of the average Einstein energy of the
phase coherent electrons is 
$<E>^{-1} = \tau_\phi/h=(\tau_0/h){\tanh}(h/2k_BT\tau_\phi)$ and
according to electron-electron theories involving large energy transfer,
$\tau_\phi = \tau_{ee} \propto L_T$\cite{altshuler}. $\alpha$ is a 
constant which only varies from 0.6 to 1.1 for all our samples.
At higher temperatures where phonons become important for dephasing,
the total phase coherence time is the inverse of  
$\tau_{\phi }^{-1} + \tau_{ep}^{-1}$.  
The solid lines drawn through {\it all}  our Au $\tau_\phi(T)$ data 
displayed in Figs. 1-3 are excellent fits to Eq. (1) including phonons.

\par

If Eq. (1) truly describes the temperature dependence of the phase 
coherence time for our samples, it should apply to all 1D and 2D
mesoscopic systems fabricated from metals and semiconductors. Fig. 4 
displays some representative examples of the previously observed 
saturation behavior of $\tau_\phi$ in a variety of 1D and 2D systems 
where the saturation temperature varies from 20 mK to 10 K. For 2D 
Au\cite{bergmann:au} and AuPd \cite{giordano2:aupd} experiments, 
Eq. (1) fits the reported phase coherence time extremely well with the 
constant $\alpha$ reduced only by a factor of $\pi$ due to the change 
in dimensionality\cite{altshuler,altshuler:book}. Similarly, for low
mobility 1D Si-MOSFETs\cite{pepper2:si} and for high mobility 1D GaAs
heterostructures\cite{hiramoto1:gaas} we find that Eq. (1) must be 
modified by replacing $L_T$ with the actual width $w$ of the  
\begin{figure}
 \vbox to 7cm {\vss\hbox to 7cm
 {\hss\
   {\includegraphics{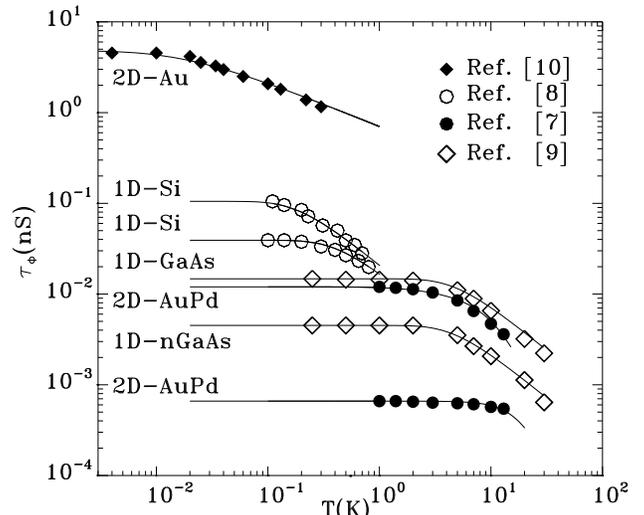}
   }
  \hss}
               }
\caption{Temperature dependence of $\tau_\phi$
in a 2D-Au, 1D-Si, 1D-GaAs and 2D-AuPd experiments. The
solid lines are our fits using Eq. (1). }

\end{figure}
\noindent 
sample and reducing $\alpha$ by $\pi^2$. The substitution of $w$ for $L_T$
is generally made because the thermal diffusion length is much larger
than the sample width. Amazingly, as shown in Table 1, the fits to
Eq. (1) describe all the published data extremely well including the
high temperature part of the data and gives us further confidence
that this zero-point dephasing mechanism correctly
describes the essential
physics of the phase coherence time for all mesoscopic samples.

\par

It is now possible to calculate the value of $\tau_0$ contained in 
Eq. (1) using a combination of previous approaches. Starting from the
fluctuation-dissipation theorem and integrating the fluctuations in the 
low temperature limit of Einstein energy over frequency we can obtain a
single effective, sample dependent, average energy. This energy can then 
be related to the dephasing time\cite{chakravarty,stern}. In a separate
publication we will describe in detail our approach to solving this 
problem\cite{next} but a preliminary result(valid only for diffusive
1D systems) which describes many of the 1D experiments published to date,
including ours, is
\begin{equation}
\tau_0 = \Big({{4\pi\hbar^2 L}\over {e^2 d^2 R m^{\ast}D^{3/2}}}\Big)^2
\end{equation}
where $R/L$ is the resistance per length, and $m^\ast$ is the effective 
mass of the electron. The calculated values $\tau_0^{calc}$ for all 1D 
experiments are listed in Table 1, and are within a factor of 2 or 3  
from the measured value $\tau_0^{exp}$. For 1D systems fabricated from 
2DEG with $L\sim L_\phi$, it is interesting to note that Eq. (2) reduces
to $(4h/m^{\ast}D)(\hbar/\Delta)$ where $\Delta$ is the separation between
energy levels in the phase coherent volume of the system. We find the 
agreement of Eq. (2) with the measured values quite remarkable considering 
the extreme sensitivity of $\tau_0$ on $D$ and $R/L$. An error of only
$25\%$ in the width and thickness of a sample can easily propagate to give 
an error of a factor of two in $\tau_0$.

\par

\begin{figure}
 \vbox to 7.2cm {\vss\hbox to 14cm
 {\hss\
   {\includegraphics{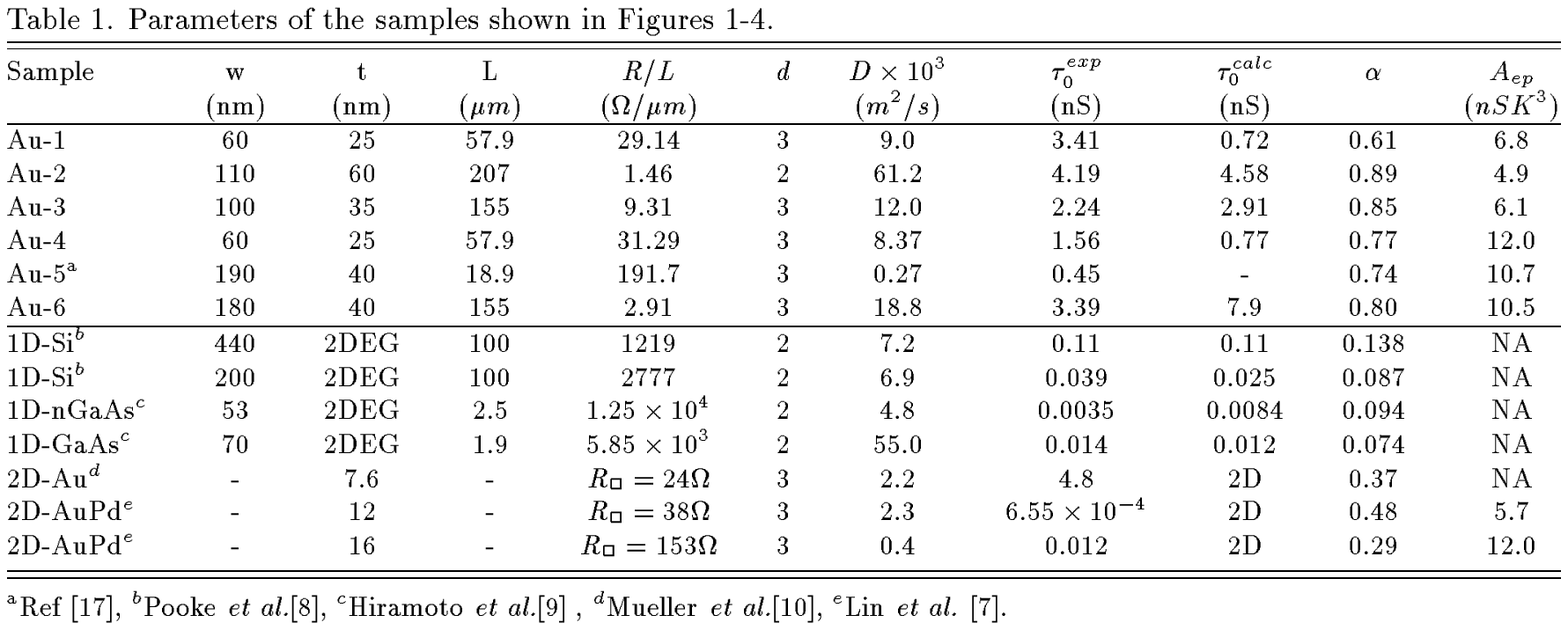}
   }
  \hss}
 }
\end{figure}
         
\par
In conclusion, we report a comprehensive set of experiments in 1D gold
wires, which clearly shows that the phase coherence time saturates at a
finite temperature. This saturation is not due to
heating, magnetic impurities, or external environmental effects.
We suggest
that zero-point fluctuations are responsible for this
observed saturation and introduce both a functional form for the
temperature dependence and a calculation for the saturation value
for the phase coherence time. We find that all our data as well as 
the data from many other groups on a wide variety of systems including
1D wires (both semiconductors and metals), and 2D films can be fit 
quite well to the form given in Eq. (1) essentially with only one  
adjustable parameter, $\tau_0$, the zero temperature phase coherence time. 
Our approach allows us, for the first time, to theoretically calculate 
$\tau_0$ for any 1D mesoscopic system from only $R/L$ and $D$, and the
agreement between the calculated value  and measured value of $\tau_0$
is extremely good in our as well as others' experiments.           

\par 

We thank G. Bergmann, N. Giordano, T. Hiramoto and M. Pepper for 
permission to use their data shown in Fig. 4. We also thank 
C. Van Haesendonck, B. L. Hu, A. Raval and S. Washburn for helpful 
discussions. This work is supported by the NSF under  
contract No. DMR9510416.

\narrowtext


\begin{references}
\bibitem{abm11}
	The sample Au-5 appears to be 2D over a part of the temperature range, 
        where Eq. (2) is not valid.
\end{references}
\end{document}